\documentclass[twocolumn, prl, aps, superscriptaddress,shortbibliography,showpacs,amsmath,amssymb,floatfix]{revtex4-1}
\usepackage{graphicx}
\usepackage{epstopdf}
\usepackage{amssymb}
\usepackage{amsmath}
\usepackage{epsfig}
\usepackage{color}
\usepackage[colorlinks,linkcolor=blue,anchorcolor=blue,citecolor=blue,urlcolor=blue]{hyperref}
\usepackage{float}

\begin{document}

\title{Dynamical instability towards finite-momentum pairing in quenched BCS superconducting phases}
\author{Beibing Huang}
\email{hbb4236@ycit.edu.cn}
\affiliation{Department of Physics, Yancheng Institute of Technology, Yancheng, 224051, P. R. China}
\author{Xiaosen Yang}
\affiliation{Department of physics, Jiangsu University, Zhenjiang, 212013, P. R. China}
\author{Ning Xu}
\affiliation{Department of Physics, Yancheng Institute of Technology, Yancheng, 224051, P. R. China}
\author{Jing Zhou}
\affiliation{Department of Science, Chongqing University of Posts and Telecommunications, Chongqing, 400065, P. R. China }
\affiliation{Beijing National Laboratory for Condensed Matter Physics, Institute of Physics, Chinese Academy of Sciences, Beijing, 100190, P. R. China}
\author{Ming Gong}
\email{gongm@ustc.edu.cn}
\affiliation{CAS Key Laboratory of Quantum Information, University of Science and Technology of China, Hefei, 230026, China}
\affiliation{Synergetic Innovation Center of Quantum Information and Quantum Physics, University of Science and Technology of China, Hefei, Anhui 230026, China}
\affiliation{CAS Center For Excellence in Quantum Information and Quantum Physics}
\date{\today}

\begin{abstract}
In this work we numerically investigate the fate of the Bardeen-Cooper-Schrieffer (BCS) pairing in the presence of quenched phase under Peierls substitution using time-dependent real
space and momentum space Bogoliubov-de Gennes equation methods and Anderson pseudospin representation method. This kind of phase imprint can be realized by modulating electric field
in ultracold atoms and illumining of THz optical pump pulse in solids with conventional and unconventional superconductors. In the case of weak phase imprint, the BCS
pairing is stable; while in the strong phase imprint, instability towards finite-momentum pairing is allowed, in which the real space and momentum space methods yield different results.
In the pulsed gauge potential, we find that this instability will not happen even with much stronger vector potential. We also show that the uniform and staggered gauge potentials yield
different behaviors. While the staggered potential induces transition from the BCS pairing to over-damped phase, the uniform gauge may enhance the pairing and will not induce to the over-damped
phase. These result may shade light on the realization of finite momentum pairing, such as  Fulde-Ferrell-Larkin-Ovchinnikov phase with dynamical modulation.
\end{abstract}

\maketitle

The quenched dynamics of Bardeen-Cooper-Schrieffer (BCS) state has attracted much interests over the past decades
\cite{kopnin2001theory, langenberg1986nonequilibrium, volkov1974collisionless, barlas2013amplitude, peronaci2015transient, krull2016coupling,
andreev2004nonequilibrium, barankov2004collective,barankov2006synchronization, yuzbashyan2006dynamical, yuzbashyan2006relaxation, yuzbashyan2005nonequilibrium,
bulgac2009large, yuzbashyan2015quantum, szymanska2005dynamics}. After quench, the asymptotic limit of this state in the long-time dynamics can exhibit several different behaviors.
During the quench of interaction strength, which may be realized in ultracold atoms by Feshbach resonance
\cite{feshbach1958unified, chin2010feshbach}, three different phases can be identified: damped oscillation phase\cite{volkov1974collisionless, yuzbashyan2006dynamical,
yuzbashyan2006relaxation}, persistent oscillation phase \cite{barankov2004collective, barankov2006synchronization} and over-damped phase \cite{barankov2004collective,
barankov2006synchronization, yuzbashyan2006dynamical}. These phases can also be seen in the spin-orbit coupling model by quenching of Zeeman field\cite{dong2015dynamical}. Recently, a new kind
of quench protocol was discussed in a series of optical pump-probe experiments in conventional and unconventional superconductors\cite{sooryakumar1980raman, measson2014amplitude, matsunaga2012nonequilibrium,
matsunaga2013higgs, matsunaga2014light, katsumi2018higgs}. Via the gauge potential ${\bf A}(t)$ to the Hamiltonian by Peierls substitution, it will enter the generalized momentum,
which may lead to new dynamics. In Ref. \onlinecite{chou2017twisting}, it was shown that only the damped oscillation and the over-damped phases are observable.

Theoretically, quenched dynamics of the BCS state is generally based on the Anderson pseudospin representation (APR) method\cite{anderson1958random}. For the fermion creation
(annihilation) operators: $c_{{\bf k}\sigma}^\dagger$ ($c_{{\bf k}\sigma}$), where ${\bf k}$ is the momentum and $\sigma = \uparrow, \downarrow$ is the spin, one can define spin operators
as $s_{\bf k}^z = {1\over 2}(c_{{\bf k}\uparrow}^\dagger c_{{\bf k}\uparrow} + c_{{\bf -k}\uparrow}^\dagger c_{{\bf -k}\uparrow}  - 1)$, $s_{\bf k}^+ =
c_{{\bf k}\uparrow}^\dagger c_{{\bf -k}\downarrow}^\dagger$, $s_{\bf k}^- =
c_{-{\bf k}\downarrow} c_{{\bf k}\uparrow}$, one finds that spin operators satisfy the standard SU(2) symmetry: $[s_{\bf k}^+, s_{\bf k}^-] = 2s_{\bf k}^z$,
$[s_{\bf k}^z, s_{\bf k}^\pm ] = \pm s_{\bf k}^\pm$. The BCS Hamiltonian can be expressed exactly using these operators in case that the kinetic energy term has inversion symmetry,
that is, $\epsilon_{{\bf k}\uparrow} = \epsilon_{-{\bf k}\downarrow}$ for spinful fermions and $\epsilon_{{\bf k}} = \epsilon_{-{\bf k}}$ for spinless fermions.
This method was widely applied to study the quench dynamics of BCS phases to identify the above mentioned long-time
behaviors\cite{barlas2013amplitude, peronaci2015transient, krull2016coupling, andreev2004nonequilibrium, barankov2004collective, barankov2006synchronization, yuzbashyan2006dynamical, yuzbashyan2006relaxation, yuzbashyan2006dynamical, barankov2004collective, yuzbashyan2005nonequilibrium, yuzbashyan2015quantum, szymanska2005dynamics, murotani2017theory}.
It may also be applied to study the dynamics of these BCS pairing with imprinted gauge potential. Some works have pointed out the possible stability of these phases toward to spatial
fluctuation in the interaction quenching \cite{chern2018nonequilibrium, dzero2009cooper}. This problem will become more serious in the pump-probe experiments in which the gauge potential directly
breaks the inversion symmetry, in which the APR method does not work any more. We ask the general question that whether the BCS pairing is still stable in the phase quench experiments.

In this work, we report that during the quench of phase the BCS pairing may become unstable towards finite-momentum pairing. To account for this mechanism, we employ the
time-dependent real-space Bogoliubov-de Gennes (TRBdG) equation, which avoids the assumption of uniform pairing in momentum space. We have also considered the fate of these phases under
BCS framework using time-dependent momentum space BdG (TKBdG) formalism. In case when APR is applicable, we will compare these results with the APR method. These results show that
under weak quenched phase the BCS pairing is stable, while in strong quenched phase, some finite-momentum pairing will be invoked. In case of pulsed gauge potential, the BCS
pairing is always very stable and finite momentum pairing is hard to be realized even with extremely strong quenched phase. These results are stimulating for the possible realization
of Fulde-Ferrell-Larkin-Ovchinnikov (FFLO) phase\cite{PhysRev.135.A550, Larkin1965INHOMOGENEOUS, RevModPhys.76.263} with dynamical modulation.

We consider the following Fermi-Hubbard model
\begin{eqnarray}
	H=-J\sum_{j\sigma} e^{i\theta_\sigma(t)}c_{j\sigma}^{\dag}c_{j+1\sigma} + \text{h.c.}-U n_{j\uparrow} n_{j\downarrow} + \mu n_j,
\label{eq-H}
\end{eqnarray}
where $c_{j\sigma}$ denotes the fermion annihilator operator at $j$-th site with spin $\sigma=\uparrow, \downarrow$ and $\mu$ is the chemical
potential. For attractive interaction, $U > 0$. An important feature is the quenched phase $e^{i\theta_\sigma(t)}$ from Peierls substitution, which may be realized using a number of methods in
different systems (see below). We will consider two different vector potentials: (I) uniform vector potential with $\theta_\sigma = \theta$; and (II) staggered vector potential with
$\theta_\uparrow = -\theta_\downarrow = \theta$. In the latter case, the Hamiltonian still respects the inversion symmetry, thus the APR method is still applicable, while the first
one does not. Our motivation is as following. When $\theta_\sigma(t) = 0$, the ground state is in BCS phase with uniform pairing, $\Delta =U \langle c_{j\downarrow}c_{j\uparrow}\rangle$.
In case (I), the ground state can be obtained from the wave function with $\theta_\sigma = 0$ via a transformation, $c_{j\sigma} \rightarrow c_{j\sigma}e^{i\theta(t)j}$,
in which the pairing may carry a finite momentum. Thus it is intriguing to explore the possible transition from BCS pairing to finite momentum pairing. In case (II) the ground state is
still in BCS phase, which can be well described by APR method. The real physics may become more complicated due to the presence of self-consistent calculation. The consequences of these
vector potentials will also be compared carefully.

We will consider two different quench protocols, which are closely related to that in experiments. To account for the finite-momentum pairing, we consider the dynamics in real space, which is termed as TRBdG.
The on-site pairing is introduced to the Hamiltonian via $Uc_{j\uparrow}^{\dag}c_{j\downarrow}^{\dag}c_{j\downarrow}c_{j\uparrow}\approx \Delta_j c_{j\uparrow}^{\dag}c_{j\downarrow}^{\dag}+ \Delta_j^*c_{j\downarrow}c_{j\uparrow}-|\Delta_j|^2/U$, and in the Nambu basis $\Phi=(c_{1\uparrow},\cdots c_{N\uparrow}, c_{1\downarrow}^{\dag}, \cdots c_{N\downarrow}^{\dag})^T$ the Hamiltonian becomes
$H = \Phi^{\dag}H_{\text{BdG}}\Phi-N\mu+\sum_j|\Delta_j|^2/U$ with
\begin{eqnarray}
	H_{\text{BdG}}=\left(\begin{array}{cc}
\mathcal{H} & \Delta \\
\Delta^* & -\mathcal{H} ^*
\end{array}\right),
\end{eqnarray}
where $\mathcal{H}_{mn}=-\mu\delta_{mn}-Je^{i\theta}(\delta_{mn+1}+\delta_{m1}\delta_{nN}) -Je^{-i\theta}(\delta_{m+1n}+\delta_{mN}\delta_{n1})$
and $\Delta_{mn}=-\Delta_m\delta_{mn}$ with $N$ being the total lattice sites with periodic boundary condition. This model is diagonalized by unitary transformation
$V = (u_1, u_2, \cdots, u_{2N})$, where $Hu_i = \epsilon_i u_i$ with $\epsilon_i$ being arranged in increasing order.

\begin{figure}
\centering
	\includegraphics[width=0.49\textwidth]{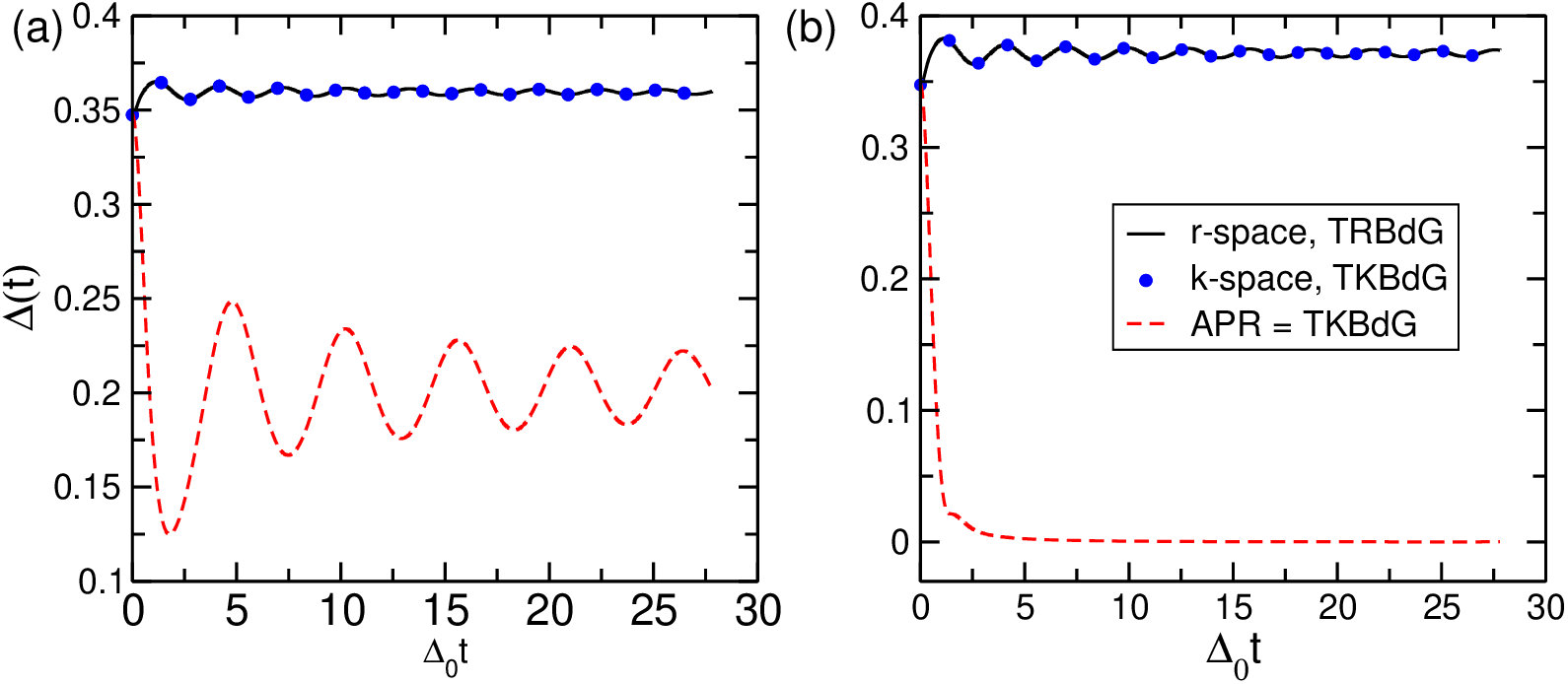}
\caption{The dynamical evolution of order parameters with $\theta_\sigma =\theta = 0.186$ (a) and 0.264 (b) based on real-space calculation (see solid line)
and momentum-space calculation (see solid symbols) with uniform BCS pairing. For comparison, we also compare with the corresponding results with $\theta_\uparrow =  - \theta_\downarrow =
	0.186$ and 0.264, respectively, which can be computed using APR method (see dashed line), which are identical to that by TKBdG (denoted as APR = TKBdG throughout this work).
	In (a), $(\bar{\Delta}, \omega) = (0.360, 0.720)$ for TRBdG and (0.203, 0.403) for APR; while
	in (b) are (0.372,0.744) for TRBdG, indicative of Higgs modes.}
	\label{fig-fig1}
\end{figure}

The coherent dynamics of this model $H$ can be solved as follows\cite{dong2015dynamical}. Firstly, $t = 0$, $H$ is diagonalized by the transformation $ \Phi=V \Gamma$ with quasi-particle operators
$\Gamma=(\gamma_{1\uparrow}, \cdots, \gamma_{N\uparrow}, \gamma_{1\downarrow}^{\dag}, \cdots  \gamma_{N\downarrow}^{\dag})^T$. The order parameter in each site is
updated by $\Delta_j=U\sum_{l=N+1}^{2N}u^*_{j+Nl}u_{jl}$, and $\Delta_j=\Delta_0$ corresponds to uniform phase. At $t>0$, the wave function is described by
\begin{eqnarray}
	i\frac{\partial}{\partial t}u_j(t)=H_{\text{BdG}}(\Delta(t))u_j(t)
	\label{eq-schrodingereq},
\end{eqnarray}
where $\Delta(t)$ is updated instantaneously by including the contribution of all eigenvectors. In some literature, the dynamics
of these state is computed by considering the evolution in Heisenberg picture\cite{chern2018nonequilibrium, PhysRevA.91.043630, PhysRevA.96.033618}, which is equivalent to the above theory.

The above real space is particularly suitable to investigate the dynamics with non-uniform pairings. In the case of uniform pairing, one may also perform the similar
calculation in momentum space; see for example Ref. \onlinecite{dong2015dynamical}, by assuming a time-dependent uniform pairing $\Delta(t)$.
This method will be termed as TKBdG, in which the update of $\Delta(t)$ is the same as that in Eq. \ref{eq-schrodingereq}.
In the case when the kinetic energy term respects the inversion symmetry with staggered gauge potential, we also perform the same calculation using APR
method in momentum space. In the above two methods, the TKBdG can be applied to all BCS phases, thus has much broader applicabilities. The APR
method, however, is more illustrative for its novel pictures in dynamics.

In following, we will present our results based on lattice up to $N=350$, which is long enough to exclude the finite size effect for the time-domain we have
considered. We have also studied the same physics in the $20\times 20$ two-dimensional lattice, which yields results qualitatively the same as that in one-dimension.
In following, without loss of generality, we consider $J=1.0$, $U/J=2.0$, $\mu/J=-0.4$. This set of parameters corresponds to a filling factor $n = 0.87$ and $\Delta_0/J = 0.35$.
The above three methods will be used simultaneously to unveil the instability of BEC phase. For convenience, we denote $\bar{\Delta}$ the mean value of the order parameter during dynamics and $\omega$ the corresponding oscillating frequency.

{\it Uniform Phase Quench}. We consider the following phase imprint protocol,
\begin{eqnarray}
\theta(t)=\theta\Theta(t)
\label{eq-model1}
\end{eqnarray}
with $\Theta(t)$ being the Heaviside step function. This pulse can be obtained in the cold atom experiments by manipulating the one-dimensional optical lattice potential $V(x,t)=V_0\cos^2{(kx+\varphi(t))}$, where the
relative phase $\varphi(t)$ between incidence and retroreflected lasers can be modulated by controlling the movement of reflection mirror\cite{morsch2006dynamics, zheng2014floquet}. In the comoving frame, one obtains a time-dependent vector potential $A(t)=-\frac{m}{
ek}\frac{d\varphi(t)}{dt}$, where $m$ is the mass of fermion atoms and $e$ is elementary charge\cite{zheng2014floquet}. Depending on the manner in which we control, $\varphi(t)$ can show many
different behaviors. If we periodically shake the reflection mirror along the $x$ direction, the relative phase will be a periodic function of time $t$. This periodic
driving has become the research hotspot in the condensed matter physics\cite{oka2009photovoltaic,
inoue2010photoinduced, lindner2011floquet, chen2018floquet, yan2017floquet, yan2016tunable, chan2016chiral} and cold atom
physics\cite{eckardt2017colloquium, goldman2014periodically, goldman2015periodically, d2014long, zhang2015chiral, struck2012tunable, struck2011quantum, jotzu2014experimental, tarruell2012creating, zheng2016fulde} for realization of exotic phases. In order to attain the Peierls phase in Eq. \ref{eq-model1}, we can displace the reflected mirror along $x$ direction in a uniform speed $v$.
As a result $\varphi(t)=k(x_0+vt)$ with $x_0$ being the initial position of mirror and $A(t)=-(mv/e)\Theta(t)$. This vector potential corresponds to a pulsed electric field. In the ultracold atoms, the
lattice constant $a\sim 10^2$nm and mass $m\sim 10^{-27}$Kg (for $^{87}$Rb), $v \sim 1$ m/s, we estimate $\theta=ea A/\hbar=a m v /\hbar \sim \mathcal{O}(1)$.

\begin{figure}
\centering
\includegraphics[width=0.49\textwidth]{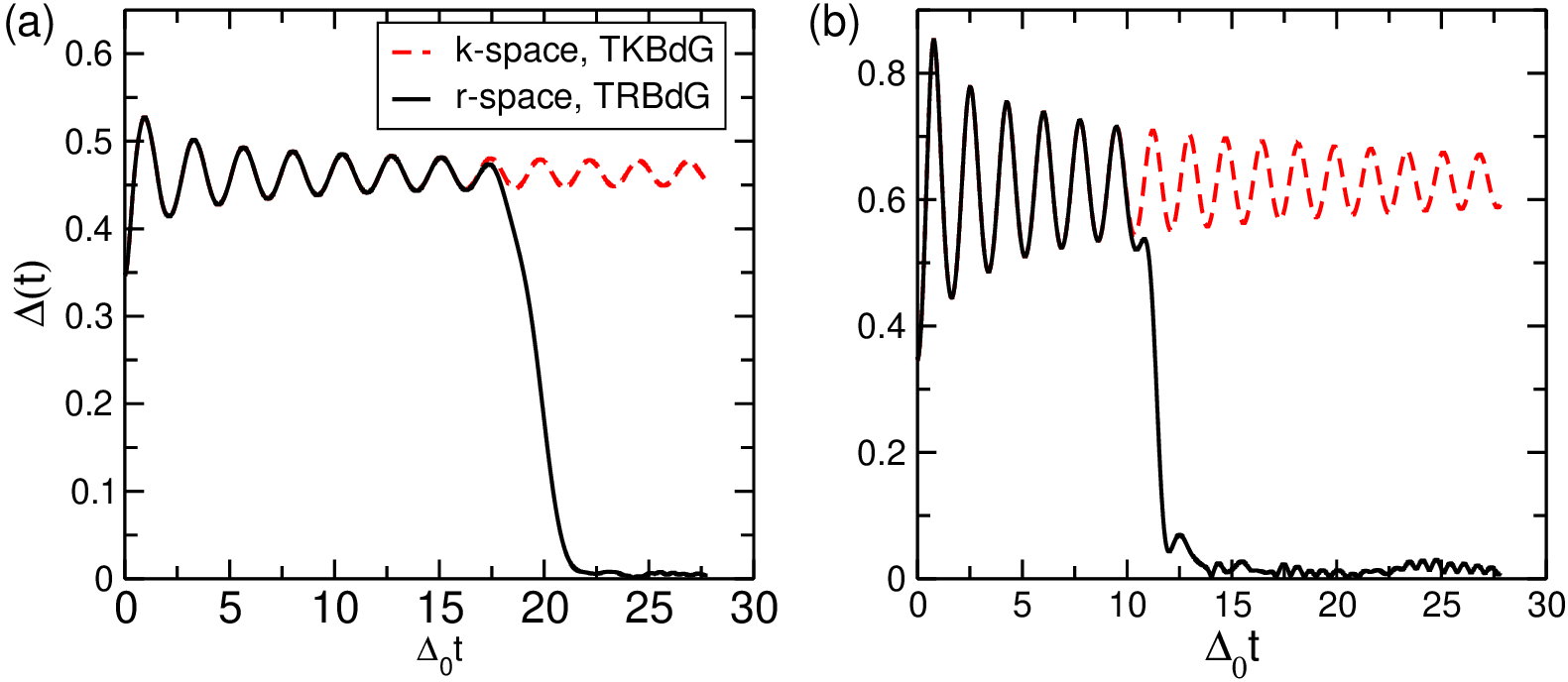}
\caption{The dynamical evolution of order parameters of model with $\theta_{\sigma} = 0.59$ (a) and 1.02 (b). The dashed line and solid line are obtained by momentum space and real-space
calculations, respectively. The same results have been obtained with a shorter or longer lattice system. The results in APR is not shown due to the over-damped phase with uniform BCS pairing in
Fig. \ref{fig-fig1}b. In (a), $(\bar{\Delta}, \omega) = (0.463, 0.926)$ and (b) is (0.625, 1.258), indicative of Higgs modes.}
\label{fig-fig2}
\end{figure}

We first consider the case with small phase quench. Our numerical results for the two phase quenches (I and II) are presented in Fig. \ref{fig-fig1}. For the uniform quench, we
find that the TRBdG and TKBdG yield the same dynamics. In this case, the order parameter is always uniform across the whole lattice. The weak damping corresponds to the damped
oscillating phase. We notice that in both cases with $\theta = 0.186$ and $0.264$, the pairing strength is enhanced and we always found $|\Delta(t)| > |\Delta_0|$ for $t > 0$.  Thus we find that the BCS phase is
stable against weak inversion symmetry breaking during quench dynamics.
This is different from the quenched phase with staggered vector potential (II), in which with very weak phase quench (Fig. \ref{fig-fig1}a), the system is in the damped oscillation phase
with $|\Delta(t)| < |\Delta_0|$ for $t > 0$. It seems that in this quenched potential the order parameter is rather sensitive to amplitude of the quenched phase. By increasing the
phase slightly from Fig. \ref{fig-fig1}a to b, the damped oscillation phase is changed to the over-damped phase with uniform BCS pairing. In the damped oscillating phase, the different magnitudes of order parameters determine different oscillating frequencies\cite{volkov1974collisionless, yuzbashyan2006dynamical, yuzbashyan2006relaxation}. In this figure, we find
\begin{equation}
	\omega = 2 \bar{\Delta},
\end{equation}
indicative of the collective amplitude modes (Higgs mode)\cite{volkov1974collisionless, yuzbashyan2006dynamical, yuzbashyan2006relaxation}. This conclusion is true in all figures for both quenched phases.

\begin{figure}
\centering
\includegraphics[width=0.49\textwidth]{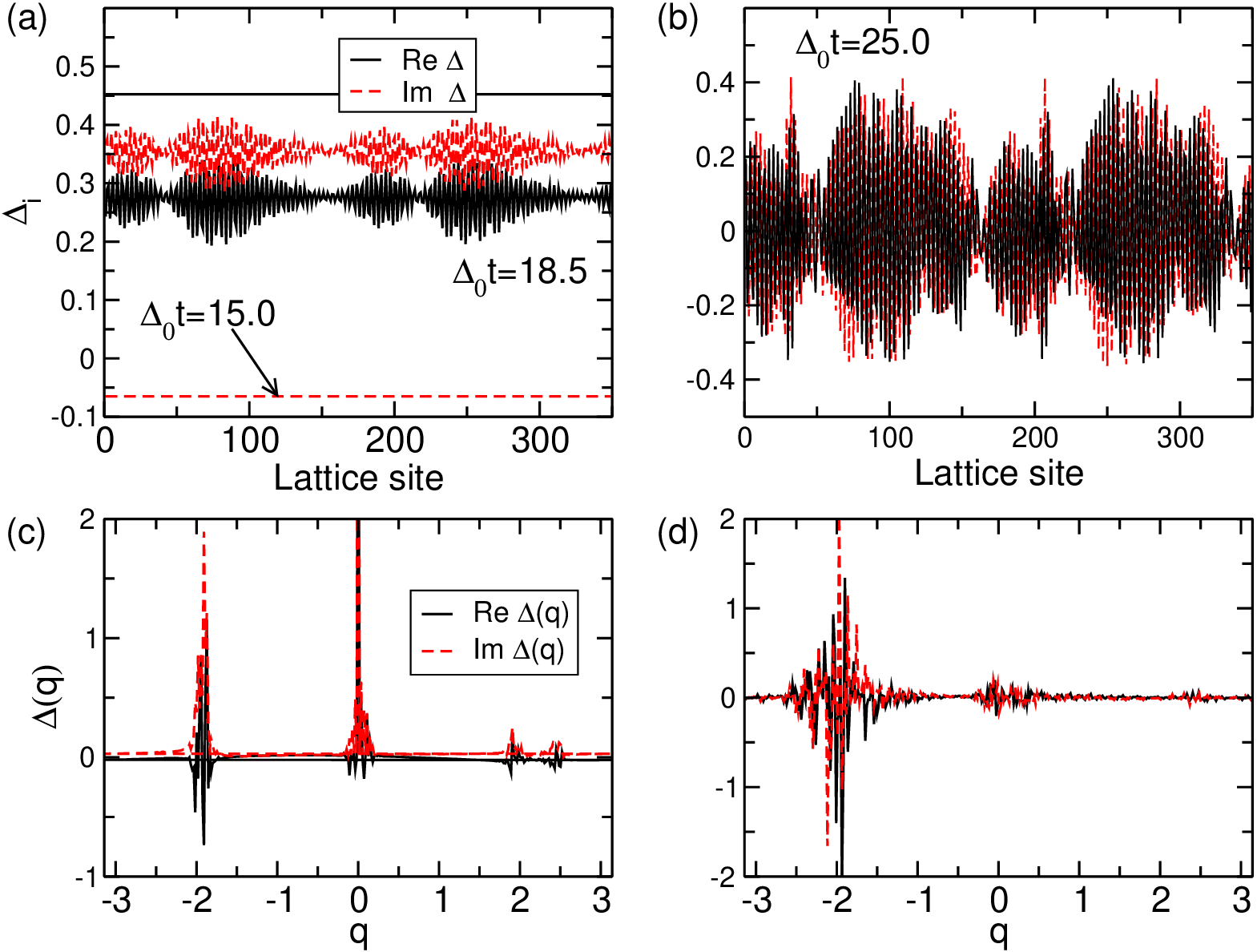}
	\caption{The profile of order parameters in real space at $\Delta_0t=15.0$ for BCS uniform pairing and 18.5 for finite-momentum pairing in (a) and $\Delta_0t=25.0$
	for finite-momentum pairing in (b), for the case considered in Fig. \ref{fig-fig2}a. The upper two panels show the real part (solid line) and imaginary part (solid line) of $\Delta_i$,
	while (c) and (d) show the corresponding pairing in momentum space. In (c), the peak at $q=0$ is truncated to signify the finite momentum pairings.}
\label{fig-fig3}
\end{figure}

With the further increasing of quenched phase, the system enters the instability regime. The results are presented in Fig. \ref{fig-fig2}. The dynamics of this
phase exhibits two different behaviors. Firstly, it exhibits damped oscillation with uniform spatial pairing. In the TRBdG method, we find that the mean order parameter $\Delta(t) =
|\sum_i \Delta_i|/N$, after a certain evolution time, will suddenly decay to a very small value. Such behavior is different from the TKBdG method, in which the damped oscillating phase
can persist for a long time. Before the instability  points, these two methods yield the same result; after that, this mean pairing strength will
quickly decreases to a small magnitude. The reason for this decrease comes not from the over-damped phase, but from the non-uniformity of pairing in real space induced by inversion
symmetry breaking, which is captured by TRBdG method (see Fig. \ref{fig-fig3}a-b). In this case, one may also compute the effective pairing in momentum space using
\begin{equation}
    \Delta_q = \langle c_{k\uparrow} c_{-k+q\downarrow}\rangle = \sum_n  e^{iqn} \langle c_{n\uparrow} c_{n\downarrow}\rangle/N.
\end{equation}
In the above formula only the on-site pairing is nonzero for $s$-wave pairing. It means that the Fourier transformation corresponds to the pairing in momentum space,
and the finite-momentum pairing is indicated by $\Delta_q \ne 0$ for $q \ne 0$. This is clearly shown in Fig. \ref{fig-fig3}c-d. Especially when the mean pairing in the whole system
is small at $t = 25.0$, the pairing in each site is still strong (see Fig. \ref{fig-fig3}b), indicating that this phase is totally different from the over-damped phase, in which the order parameter decays exponentially to zero\cite{barankov2004collective, barankov2006synchronization, yuzbashyan2006dynamical}. It is a common feature for all FFLO phases that the mean order parameter should be zero.
However, one should be noticed that the observed finite-momentum pairing is not exactly the same as the conventional FFLO pairings in literature\cite{PhysRev.135.A550, Larkin1965INHOMOGENEOUS, RevModPhys.76.263}. In our model, during the dynamical evolution, a lot of momenta $q$ are involved. It is an interesting question in future to definitely quench the BCS pairing to pure FFLO pairing.
It is necessary to address that we do not find the over-damped phase, which appears in APR for the staggered vector potential (II).

There is a fundamental reason why this instability can happen. In the standard BCS wave function (for $q = 0$), $|\Psi_q\rangle = \prod_k (u_k + v_k c_{k\uparrow}^\dagger
c_{-k+q\downarrow}^\dagger)|0\rangle$, where $|0\rangle$ is the vacuum state. The evolution of this BCS state to the finite momentum pairing requires the breaking of {\it all} weakly
bounded Cooper pairs with momentum $k$ and $-k$, and then simultaneously form new bound states between $k$ and $-k+q$, with $q\ne 0$, which is challenging. This may explain why the BCS
state is robust against weak phase imprint. However, this channel is indeed feasible if we look at this problem by writing the wave function in real
space as $|\Psi_q\rangle = \exp(\int dxdy f_{x -y} c_{x\uparrow}^\dagger c_{y\downarrow}^\dagger)|0\rangle$, where $f_{x-y}$ is related to the Fourier transformation of
$u_k/v_k$\cite{read2000paired}. The non-uniformity of pairing in real space gives rise to the instability. Next, we need to point out the novelty of this result. In previous researches, people
try to realize the long-sought finite momentum pairing by a number of different mechanisms. In the most conventional idea of FFLO, it is realized by competition between
pairing energy and magnetization energy. This mechanism is hard to be realized in experiments. In ultracold atoms, this phase was widely explored in the spin-imbalanced
systems\cite{sheehy2006bec, hu2006mean}. The spin-orbit coupled system provides a new mechanism for this phase. In this case, the finite momentum pairing can be induced by
inversion symmetry breaking\cite{zheng2013route, zheng2014fflo, wu2013unconventional, zhou2013exotic, cao2014gapless}. In the present work, we show that this phase may be
induced by dynamical quenching. This instability may be probed experimentally by pairwise projecting the fermionic pair wave function onto molecular wave function via a rapid
sweep to the strong coupling side of the Feshbach resonance. Then the technique of time-of-flight imaging can be used to measure the molecular momentum distribution, which should
reveal multiple peaks associated with pairing momenta. This experimental procedure has successfully observed the condensation of fermionic atom pairs in the weakly interacting BCS
side \cite{regal2004observation, zwierlein2004condensation}.

\begin{figure}
\centering
\includegraphics[width=0.49\textwidth]{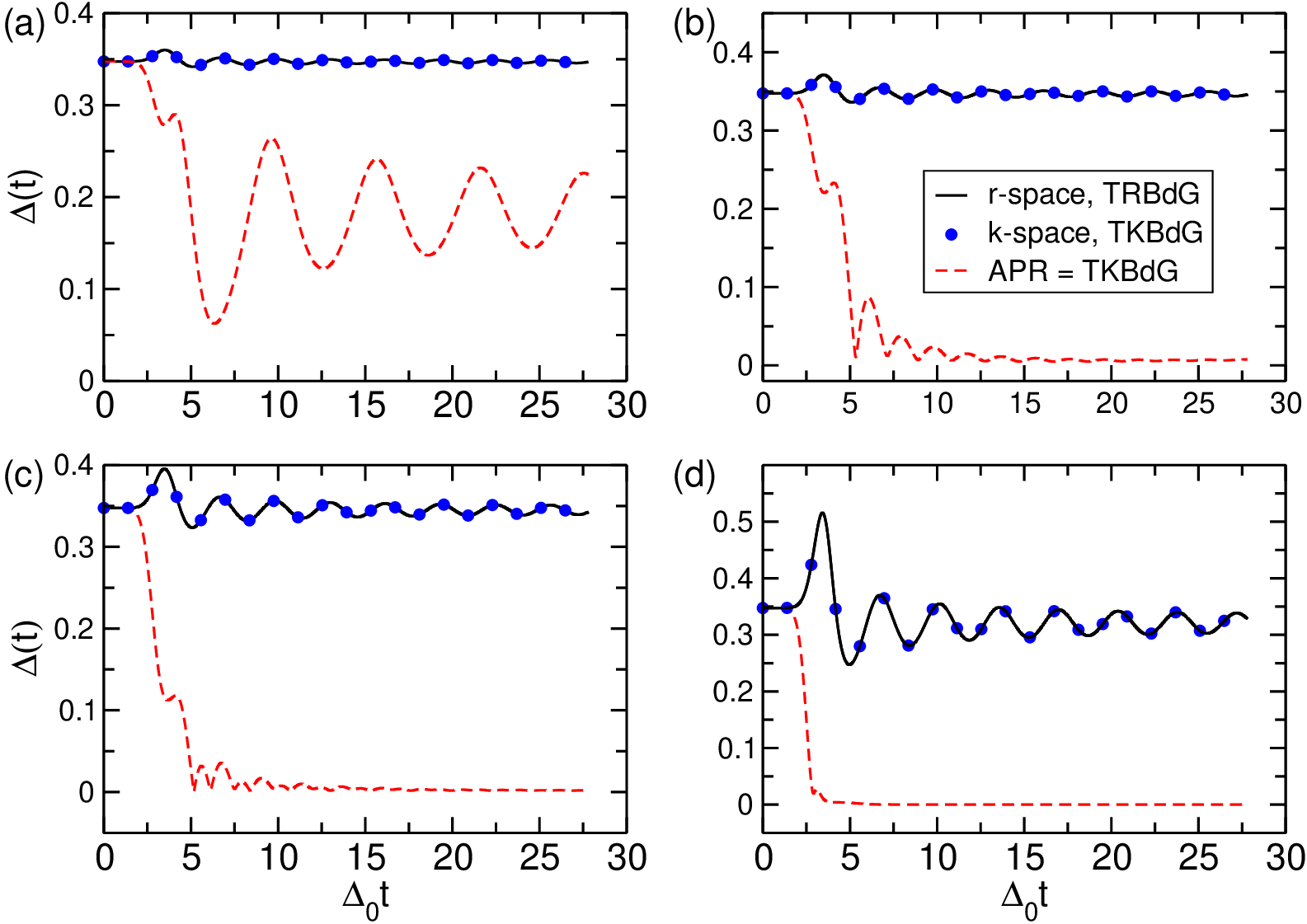}
\caption{The dynamical evolution of order parameters under uniform quench with an gaussian vector potential (see Eq. \ref{gaussian}).
The four panels corresponds to $\theta = 0.167$ , (b) $0.228$, (c) $0.323$, (d) $0.589$ with $\theta=eA_0a$. In all figures, $B_0 =0.34$. For comparison, we also present the
same phase quench using staggered vector potential (II). The values of $(\bar{\Delta},\omega)$ in (a) are (0.347,0.693) for TRBdG and (0.183, 0.365) for APR; (b) are (0.347, 0.494);
(c) are (0.345, 0.690); and (d) are (0.320, 0.640), all indicative of Higgs modes.}
\label{fig-fig4}
\end{figure}

{\it Pulsed Phase Quench}: Now we discuss a different quench protocol enlightened by pump-probe experiments
on the BCS superconductors. In a widely known experiment \cite{matsunaga2014light, matsunaga2013higgs}, an intense monocycle THz pulse
was injected into the BCS superconductor Nb${}_{1-x}$Ti${}_{x}$N;
the subsequent dynamical evolution was observed by detecting the
optical conductivity, which can be determined by the absorption or
reflection of the probe pulse. Chou et al. \cite{chou2017twisting} have
selected the following Gaussian vector potential
\begin{eqnarray}
	A(t)=A_0e^{-8(B_0t/\pi-1)^2}\label{gaussian}\Theta(t).
\end{eqnarray}
to simulate the optical pulse in this experiment, where $A_0$ is the
peak amplitude of the vector potential and $B_0$ is a tunable parameter
in experiments. Again, we consider the dynamics with vector
gauges I and II. The corresponding data are presented in
Fig. \ref{fig-fig4}. With the increasing of vector potential strength,
the APR predicts the transition from damped oscillating phase to the
over-damped phase. In the long-time limit when $A(t)$ vanishes, the
over-damped phase will not recover to the damped oscillating phase.
The dynamics are totally different based on TRBdG, in which the
weakly damped phase always presented, indicating robustness of BCS pairing.
In case of extremely strong pulse, we find that the pairing is still in the
BCS phase. Since the optical conductivity generally depends on the order parameter\cite{chou2017twisting},
these findings are very vital for properly illustrating THz optical pump-probe experiments in
superconductors.

To conclude, we investigate the robustness of the BCS phase upon quenched phase using time-dependent
real space and momentum space BdG formalism, as well as the APR method. These results show that
the BCS phase may become unstable towards finite-momentum pairing under uniform quenched phase.
In the pulsed phase, the BCS state is always stable. They are related to experiments
in ultracold atoms and pump-probe experiments in superconductors. It may provide a
new mechanism for the realization of finite-momentum pairing in FFLO phases. Our results
shade great light on the realization of this long-sought phase using dynamical modulation method
by fine tailoring the vector potentials.

\textit{Acknowledgements.} B.H.,  X.Y. and N.X. are supported by
NSFC (No. 11547047, No. 11504143, No. 11404278). J.Z. is supported
by NSFC under Grant No. 11504038, Chongqing Fundamental, Frontier
Research Program (Grant No. cstc2015jcyjA00013), and Foundation of
Education Committees of Chongqing (Grant No. KJ1500411).
M.G. is supported by the National Youth Thousand Talents Program (No. KJ2030000001), the USTC start-up funding (No. KY2030000053),
the national natural science foundation (NSFC) under grant No. 11774328 and
National Key Research and Development Program of China (No. 2016YFA0301700).

\bibliography{ref}


\end{document}